\documentclass[preprint,12pt]{elsarticle}

\usepackage{amssymb}
\usepackage{graphicx}
\usepackage{float}
\usepackage{hhline}
\usepackage{hyperref}
\usepackage{bbm}
\usepackage{comment}
\usepackage{mathtools}
\usepackage{nicematrix}
\NiceMatrixOptions{transparent}

\DeclareMathOperator*{\argmax}{arg\,max}
\DeclareMathOperator*{\argmin}{arg\,min}

\usepackage{algorithm}

\usepackage{tikz}
\usepackage{tikz-cd}
\usetikzlibrary{matrix,fit}
\usetikzlibrary{calc}
\usetikzlibrary{decorations.markings}
\usetikzlibrary{arrows.meta}
\usetikzlibrary{automata, positioning}

\newtheorem{theorem}{Theorem}
\newcommand{\proofofref}{}
\newproof{zproofof}{Proof of \proofofref}
\newenvironment{proofof}[1]
 {\renewcommand{\proofofref}{#1}\zproofof}
 {\endzproofof}

\begin{document}

\begin{frontmatter}



\title{The Frequency of Convergent Games under Best-Response Dynamics}

\author{Samuel C. Wiese}
\address{Department of Computer Science, University of Oxford, Oxford OX1 3QD, UK}
\address{Institute for New Economic Thinking, University of Oxford, Oxford OX1 3UQ, UK}
\ead{samuel.wiese@wolfson.ox.ac.uk}

\author{Torsten Heinrich}
\address{Department of Economics and Business Administration, Chemnitz University of Technology, 09107 Chemnitz, Germany}
\address{Institute for New Economic Thinking, University of Oxford, Oxford OX1 3UQ, UK}
\address{Oxford Martin School, University of Oxford, Oxford OX1 3BD, UK}
\ead{torsten.heinrich@wirtschaft.tu-chemnitz.de}

\begin{abstract}
Generating payoff matrices of normal-form games at random, we calculate the frequency of games with a unique pure strategy Nash equilibrium in the ensemble of $n$-player, $m$-strategy games. These are perfectly predictable as they must converge to the Nash equilibrium. We then consider a wider class of games that converge under a best-response dynamic, in which each player chooses their optimal pure strategy successively. We show that the frequency of convergent games goes to zero as the number of players or the number of strategies goes to infinity. In the $2$-player case, we show that for large games with at least $10$ strategies, convergent games with multiple pure strategy Nash equilibria are more likely than games with a unique Nash equilibrium. Our novel approach uses an $n$-partite graph to describe games.
\end{abstract}

\begin{keyword}
Pure Nash equilibrium \sep best-response dynamics \sep random games

\MSC 91A10 \sep 91A06
\end{keyword}

\end{frontmatter}

\section{Introduction}
A Nash equilibrium in a normal form game is a strategy profile such that, given the choice of the other players, no player has an incentive to make a different choice. If the Nash equilibrium is in pure strategies, we call it \emph{pure strategy Nash equilibrium} (PSNE), otherwise \emph{mixed strategy Nash equilibrium} (MSNE). John Nash showed that any game with a finite number of players and strategies has atleast one MSNE (Nash \cite{JN1, JN2}). This is not the case for PSNEs.

Consider an $n$-player, $m$-strategy normal form game and assume that players choose their optimal strategy (facing previous optimal strategies of the opponents) in a clockwork sequence -- player 1 goes first, then player 2, etc. until its player 1's turn again. We call a game \emph{convergent}, if after a sufficiently large number of turns, no player changes their strategy under the described dynamic.

We describe such games by an $n$-partite graph with each node corresponding to a pure strategy profile of the strategy choices of all but one player, and each edge corresponding to the optimal strategy choice (\emph{best response}). A PSNE corresponds to a shortest possible cycle of length $n$.

In general, there are three types of games:
\begin{itemize}
\item Type A: Convergent games with a unique PSNE
\item Type B: Convergent games with multiple PSNEs
\item Type C: Non-convergent games
\end{itemize}

Type A games (for instance, the Prisoners' Dilemma) are very easy to understand and perfectly predictable. They converge to the PSNE. As we may re-arrange the strategies of the players, Type B games are coordination games. An example for a  Type C games is Matching Pennies. Type B and Type C games have at least one MSNE.

We will investigate the likelihood of randomly created games that converge (Type A and Type B) in the ensemble of games with a given number of players and a given number of strategies available to each player. The frequencies can provide insights into predictability and stability of equilibria in economic systems. For situations that are conveniently modelled by low-dimensional (e.g. 2-player 2-strategy) games, this would be obvious. For trading behaviour in financial markets, innovation systems, or social behaviour during a crisis (say the Covid-19 pandemic), this is different.

\subsection{Related Work}
Several papers have considered aspects related to the number of PSNE in games with random payoffs. We briefly consider the papers that dealt with random payoffs that are i.i.d. from a continuous distribution.

Goldman \cite{AJG} considered zero-sum $2$-player games and showed that the probability of having a PSNE goes to zero as the number of strategies grows. Goldberg et al. \cite{KG} considered general $2$-player games and showed that the probability of having at least one PSNE converges to $1-\exp(-1)$ as the number of strategies goes to infinity. Dresher \cite{MD} generalized this result to the case of an arbitrary finite number of players. Powers \cite{IYP} showed that, when the number of strategies of at least two players goes to infinity, the distribution of the number of PSNEs converges to Poisson(1). Stanford \cite{WS95} derived an exact formula for the distribution of the number of PSNEs in random games. Stanford \cite{WS96} showed that for two-person symmetric games, the number of symmetric and asymmetric PSNEs converges to a Poisson distribution. More recently, Pangallo et al. \cite{MP} obtained exact results for the frequency of one or more PSNEs in the $2$-player case. Alon et al. \cite{ARY} studied the frequency of dominance-solvable games and obtained an exact formula for the $2$-player case. Dominance-solvable games are necessarily convergent, but not vice versa, so we study a larger class of games (containing, for instance, coordination games). The unique PSNE in Type A games are called \emph{Cournot stable}; this class of games was studied by Moulin \cite{M}.

\subsection{Our Contribution}
We introduce an $n$-partite graph describing the best responses of a game and use it to obtain the frequency of randomly created games with a unique PSNE in the ensemble of $n$-player, $m$-strategy games. These games are perfectly predictable. We then study games with more than one PSNE, that are convergent under best-response dynamics, in which each player successively chooses their optimal pure strategy. We show that convergent games with a smaller number of PSNEs are more common than convergent games with a higher number of PSNEs. We obtain an exact frequency for convergent $2$-player games with any given number of PSNEs. We finally highlight that for $2$-players and less than $10$ strategies, games with a unique PSNE are more common than convergent games with multiple PSNEs, otherwise less common.

\section{Methods}

\subsection{Notation}
A game with $n\geq 2$ players and $m\geq 2$ strategies available to each player is a tuple $(N,M,\{u_i\}_{i\in N})$ where $N=\{1,\dots,n\}$ is the set of players, $M=\{1,\dots,m\}$ the set of strategies for each player, and $u_i:M^n\rightarrow \mathbb{R}$ a payoff function. A \emph{strategy profile} $s=(s_1,\dots,s_n)\in M^n$ is a set of strategies for each player. An \emph{environment} for player $i$ is a set $s_{-i}\in M^{n-1}$ of strategies chosen by each player but $i$. A \emph{best response} $b_i$ for player $i$ is a mapping from the set of environments of $i$ to the set of non-empty subsets of $i$'s strategies and is defined by
\begin{equation*}
b_i(s_{-i}) := \argmax_{s_i\in M} u_i\left(s_i,s_{-i}\right).
\end{equation*}
A strategy profile $s\in M^n$ is a \emph{pure strategy Nash equilibrium} (PSNE) if for all $i\in N$ and all $s_i\in M$,
\begin{equation*}
u_i(s) \geq u_i(s_i,s_{-i}).
\end{equation*}
Equivalently, $s\in M^n$ is a PSNE if for all $i\in N$ and all $s_i\in M$, $s_i\in b_i(s_{-i})$. A game is \emph{non-degenerate}, if for each player $i$ and environment $s_{-i}$, the best-response $b_i(s_{-i})$ is a singleton; we then write $s_i=b_i(s_{-i})$. Similarly, a \emph{mixed strategy Nash equilibrium} (MSNE) is a strategy profile in mixed strategies.

\newcommand{\textstar}{$\star$}
\subsection{Games as Graphs}
The best-response structure of a game can be represented with a best-response digraph whose vertex set is the set of strategy profiles $M^n$ and whose edges are constructed as follows: for each $i\in N$ and each pair of distinct vertices $s=(s_i,s_{-i})$ and $s'=(s'_i,s_{-i})$, place a directed edge from $s$ to $s'$ if and only if $s'_i=b_i(s_{-i})$. There are edges only between strategy profiles that differ in exactly one coordinate. 

We now introduce an $n$-\emph{partite graph} as an additional representation of the best responses for a given fixed sequence of players. There is a total of $nm^{n-1}$ nodes in $n$ groups, each group corresponding to a player and each node corresponding to an environment of a player. At each node, a player chooses the best response; formally the edges are constructed as follows: for each pair $(i,j)$ of players, where $j$ moves directly after $i$, and each environment $s_{-i}=\left(s_j,s_{-i,-j}\right)$ (where $s_{-i,-j}$ is $s_{-i}$ without the strategy choice of $j$), place a directed edge from $s_{-i}$ to another environment $s'_{-j}=(s'_i,s_{-i,-j})$, if and only if
\begin{equation}
\label{eq:fullcond}
s'_i=b_i(s_{-i}). \tag{\textstar}
\end{equation}
As we can assume that games are non-degenerate, each node in a graph representing a game has an out-degree of $1$. A PSNE corresponds to a cycle of length $n$. Each player chooses among $m$ strategies at each node, thereby the total number of possible arrangements is $m^{nm^{n-1}}$, each equally likely.

We call the $n$-partite graph constructed as above but without the condition \eqref{eq:fullcond} the \emph{full $n$-partite graph} (see Figure \ref{fig:fc32} (left)). Any $n$-partite graph corresponding to a given game is a subgraph of the full $n$-partite graph. We will call a node \emph{free}, if its out-degree is $m$, and \emph{fixed}, if its out-degree is $1$.

\begin{figure}[H]
\centering
\begin{minipage}{\textwidth}
\centering
\normalfont
\begin{tabular}{r c r c | c | c |}
& & \textcolor{blue}{Pl. 3} & & \textcolor{blue}{V} & \textcolor{blue}{VI} \\\hline

 & \textcolor{red}{I} & Pl. 2 & III & (\textcolor{red}{0},0,\textcolor{blue}{0}) & (\textcolor{red}{0},0,\textcolor{blue}{1}) \\

\textcolor{red}{Pl. 1} & & & IV  & (\textcolor{red}{1},1,\textcolor{blue}{1}) & (\textcolor{red}{0},1,\textcolor{blue}{0}) \\

\hhline{~~--|-|-|}

& \textcolor{red}{II} & Pl. 2 & III & (\textcolor{red}{1},0,\textcolor{blue}{1}) & (\textcolor{red}{1},1,\textcolor{blue}{0}) \\

& & & IV & (\textcolor{red}{0},1,\textcolor{blue}{0}) & (\textcolor{red}{1},0,\textcolor{blue}{1}) \\                 
\hline
\end{tabular}
\vspace{0.2cm}
\end{minipage}
\begin{minipage}{0.37\textwidth}
\centering
\includegraphics[width=\linewidth]{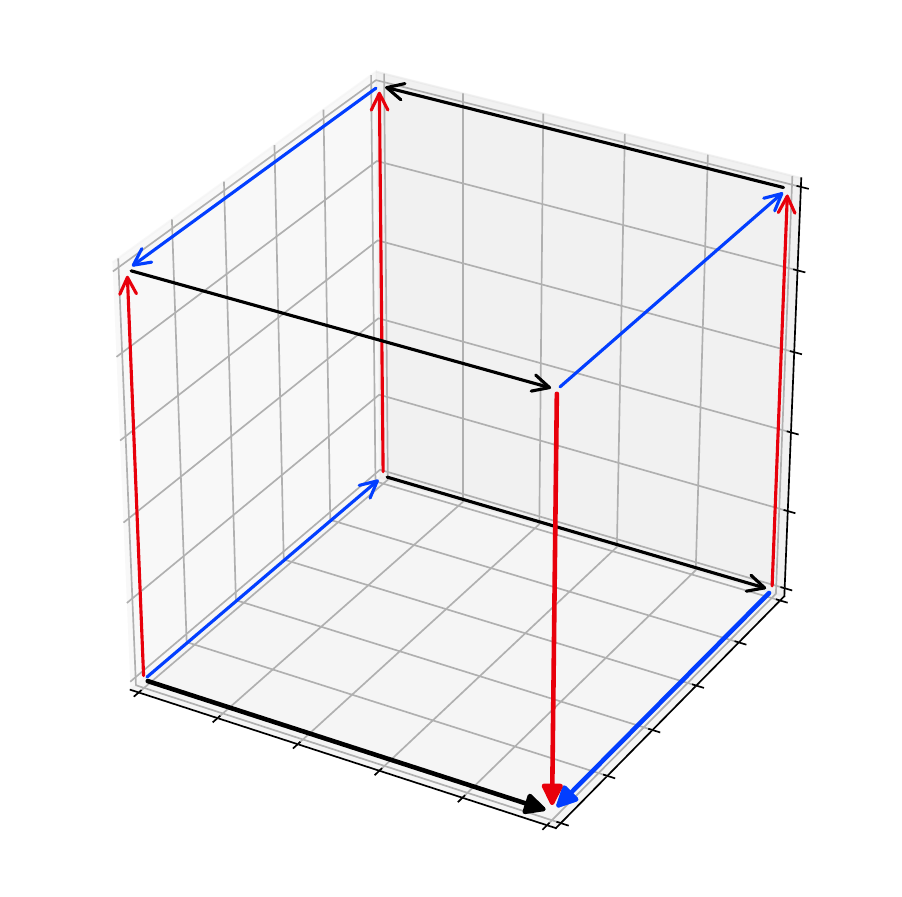}
\end{minipage}
\hfill
\begin{minipage}{0.62\textwidth}
\centering
\normalfont
\begin{tikzpicture}%
  [>=stealth,
   shorten >=1pt,
   node distance=1.1cm,
   on grid,
   auto,
   every state/.style={draw=black!60, fill=black!5, very thick, minimum width = 0.9cm}
  ]
\node[state] (A1)               {\scriptsize III-\textcolor{blue}{V}};
\node[state] (A2) [right=of A1] {\scriptsize III-\textcolor{blue}{VI}};
\node[state] (A3) [right=of A2] {\scriptsize IV-\textcolor{blue}{V}};
\node[state] (A4) [right=of A3] {\scriptsize IV-\textcolor{blue}{VI}};

\node[state] (B1) [left=1.3cm of A1] {\scriptsize \textcolor{red}{I}-\textcolor{blue}{V}};
\node[state] (B2) [below=of B1] {\scriptsize \textcolor{red}{I}-\textcolor{blue}{VI}};
\node[state] (B3) [below=of B2] {\scriptsize \textcolor{red}{II}-\textcolor{blue}{V}};
\node[state] (B4) [below=of B3] {\scriptsize \textcolor{red}{II}-\textcolor{blue}{VI}};

\node[state] (C1) [right=1.3cm of A4] {\scriptsize \textcolor{red}{I}-III};
\node[state] (C2) [below=of C1] {\scriptsize \textcolor{red}{I}-IV};
\node[state] (C3) [below=of C2] {\scriptsize \textcolor{red}{II}-III};
\node[state] (C4) [below=of C3] {\scriptsize \textcolor{red}{II}-IV};

\path[->]
(A1) edge[red, bend left=20] node {} (B3)
(A2) edge[red] node {} (B4)
(A3) edge[red, bend left=40, line width=1.2pt] node {} (B1)
(A4) edge[red, bend left=20] node {} (B4)

(B1) edge[black, bend right=40, line width=1.2pt] node {} (C2)
(B2) edge[black] node {} (C2)
(B3) edge[black] node {} (C4)
(B4) edge[black] node {} (C3)

(C1) edge[blue, bend left=40] node {} (A2)
(C2) edge[blue, bend left=40, line width=1.2pt] node {} (A3)
(C3) edge[blue, bend left=20] node {} (A1)
(C4) edge[blue, bend left=20] node {} (A4)
;
\end{tikzpicture}
\end{minipage}
\caption{A $3$-player, $2$-strategy game with one PSNE and the corresponding graph representations. The best responses corresponding to the PSNE (I-IV-V) are highlighted.}
\label{fig:brs_ex}
\end{figure}
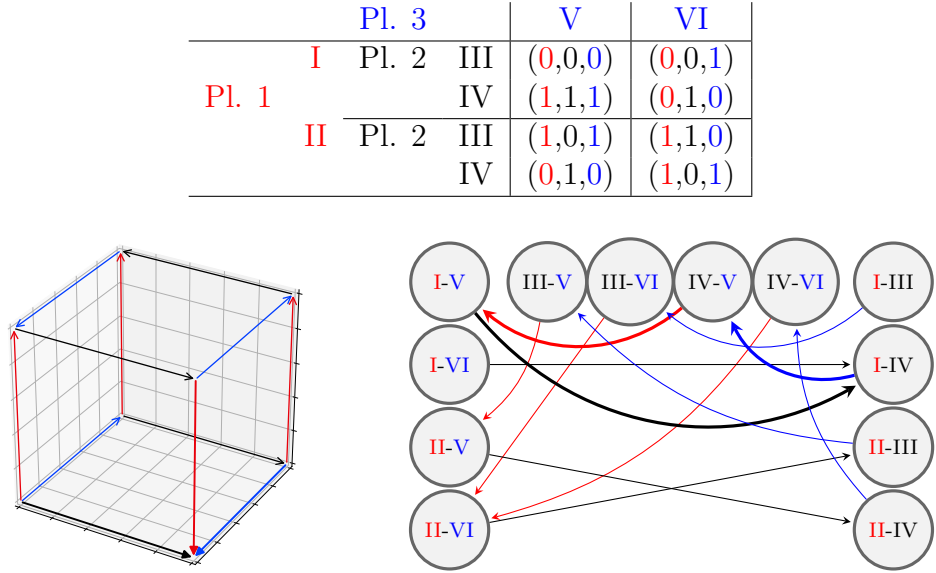

Figure \ref{fig:brs_ex} shows a $3$-player, $2$-strategy game with one PSNE. On the left is the corresponding best-response digraph, on the right the $3$-partite network with playing sequence 1-2-3.

\section{Results}

\subsection{Type A: Convergent games with a unique PSNE}
We generate $n$-player, $m$-strategy games at random by drawing $m^n$ tuples of payoffs from a multivariate normal distribution with zero mean, unit variance and identity correlation matrix. This ensures that randomly created games are almost surely non-degenerate. Let $p_{n,m}^k$ denote the frequency of $n$-player, $m$-strategy convergent games with exactly $k$ PSNEs.

\begin{theorem}
\label{maintheorem}
The frequency of games with one unique PSNE in the ensemble is given by
\begin{equation*}
p_{n,m}^1 = r^{n-1}+\frac{m-1}{m-r} \left(\left(\frac{r}{m}\right)^{n-1}-1\right)
\end{equation*}
where $r:=\frac{m-1}{m^n}+1$.
\end{theorem}

\begin{figure}[H]
\includegraphics[width=\linewidth]{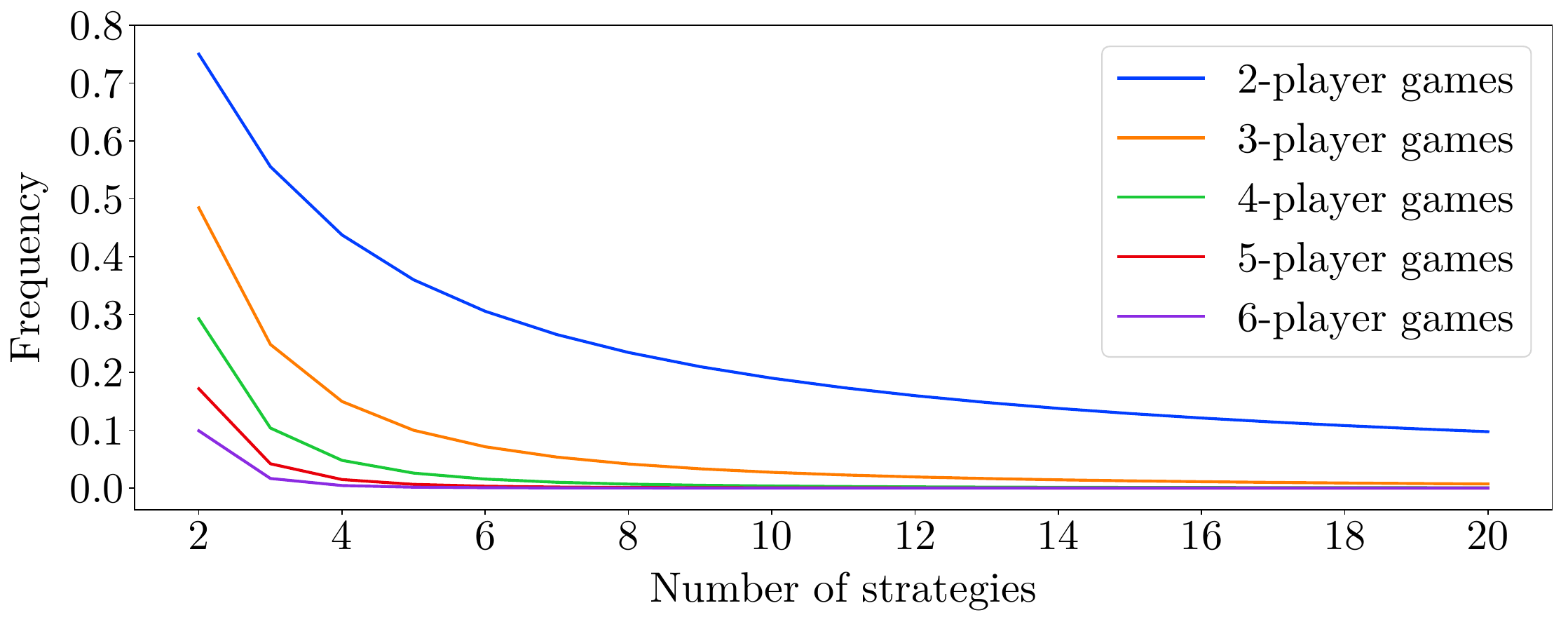}
\caption{The frequency of randomly drawn games that have a unique PSNE.}
\label{fig:freq}
\end{figure}

Note, that the frequency $p_{n,m}^1\rightarrow 0$ as the number of strategies or the number of players goes to infinity, and that $p_{n,m}^1$ is decreasing in both $n$ and $m$. For instance:
\begin{align*}
p_{2,m}^1 ={}& \frac{1}{m}\left(2-\frac{1}{m}\right) \\
p_{3,m}^1 ={}& \frac{1}{m^2}\left(3 - \frac{3}{m} + \frac{3}{m^2} - \frac{3}{m^3} + \frac{1}{m^4}\right) \\
p_{4,m}^1 ={}& \frac{1}{m^3}\left(4 - \frac{4}{m} + \frac{6}{m^3} - \frac{8}{m^4} + \frac{2}{m^5} + \frac{4}{m^6} - \frac{6}{m^7} + \frac{4}{m^8} - \frac{1}{m^9}\right) \\
p_{5,m}^1 ={}& \frac{1}{m^4} \bigg(5 - \frac{5}{m} + \frac{10}{m^4} - \frac{15}{m^5} + \frac{5}{m^6} + \frac{10}{m^8} - \frac{20}{m^9} + \frac{15}{m^{10}} - \frac{5}{m^{11}} + \frac{5}{m^{12}} -\frac{10}{m^{13}} \\ &\,\qquad + \frac{10}{m^{14}} - \frac{5}{m^{15}} + \frac{1}{m^{16}}\bigg)
\end{align*}
Figure \ref{fig:freq} shows the frequency of randomly created games with a unique PSNE.

\subsection{Type B: Convergent games with multiple PSNEs}
We can bound the frequency of convergent games with more than one PSNE from above:
\begin{theorem}
\label{upperbound}
For $k_1<k_2$, we have\ $p_{n,m}^{k_1} > p_{n,m}^{k_2}$.
\end{theorem}

This implies that for every $k$, $p^k_{n,m}\rightarrow 0$ as the number of strategies or the number of players goes to infinity. We computed for $3$-player, $2$-strategy games that $p_{3,2}^1=\frac{1984}{4096}\approx 48.43\%$, $p_{3,2}^2=\frac{828}{4096}\approx 20.21\%$, $p_{3,2}^3=\frac{56}{4096}\approx 1.37\%$, $p_{3,2}^4=\frac{2}{4096}\approx 0.049\%$.

In two-player games, we can exactly state the frequency of games with $k$ PSNEs.

\begin{theorem}
\label{twoplayers}
The frequency of $2$-player, $m$-strategy convergent games with exactly $k$ PSNEs in the ensemble is given by
\begin{equation*}
p_{2,m}^k = \frac{2m-k}{m^{2k+2} (k-1)!}  \left(\frac{m!}{(m-k)!}\right)^2.
\end{equation*}
for $k\leq m$, and is otherwise $0$.
\end{theorem}

The frequency of drawing a $2$-player convergent game (Type A or Type B) is then given by $\sum_{k=1}^m p_{2,m}^k$, the frequency of Type B games only is $\sum_{k=2}^m p_{2,m}^k$. Numerical evidence shows that Type A games are more common than Type B games for $m=2,\dots,9$, and less common for $m\geq 10$.

\begin{figure}[H]
\includegraphics[width=\linewidth]{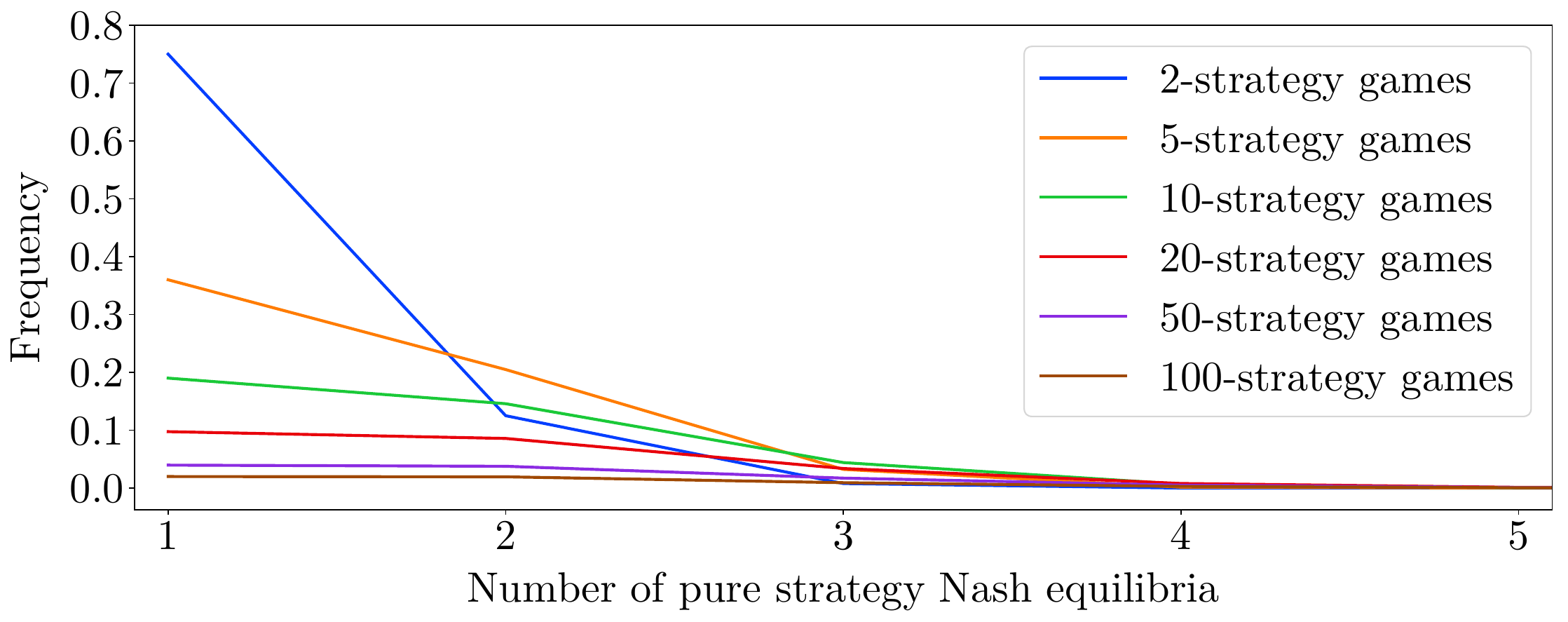}
\caption{The frequency of randomly drawn convergent $2$-player games that have a given number of PSNEs.}
\label{fig:freq_two}
\end{figure}

\begin{figure}[H]
\includegraphics[width=\linewidth]{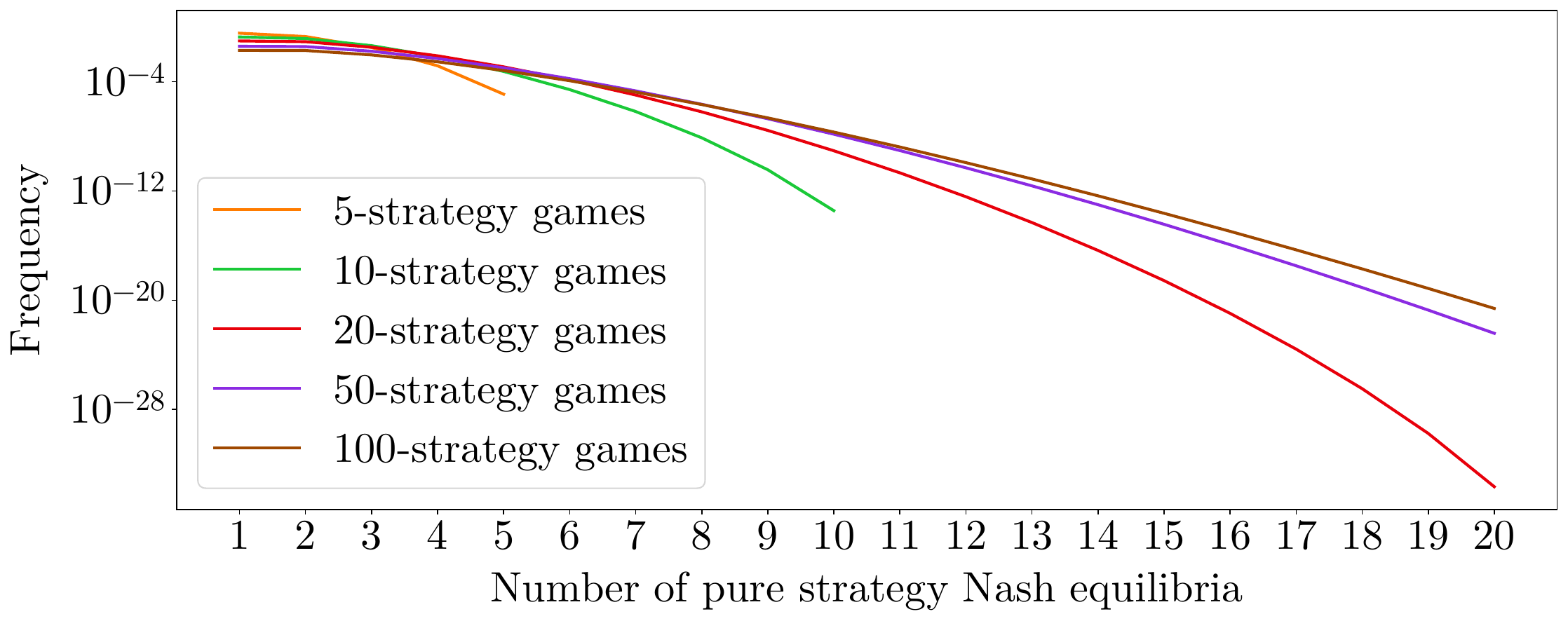}
\caption{The frequency of randomly drawn convergent $2$-player games that have a given number of PSNEs where the frequency is log-scaled.}
\label{fig:freq_two_log}
\end{figure}

Figures \ref{fig:freq_two} and \ref{fig:freq_two_log} show the frequency of randomly drawn convergent $2$-player games that have a given number of PSNEs.

\clearpage

\section{Conclusion}
We have investigated the frequency of games that are convergent under a best-response dynamic, in which each player chooses their optimal pure strategy successively. Such games may either be perfectly predictable, if they have a unique PSNE, or have multiple PSNEs. We analytically computed the frequency of the first type by using a novel graph-theoretic approach for describing games, and showed that if we let the number of players or the number of strategies go to infinity, almost all games do not converge. We also showed that games with a higher number of PSNEs are less common than games with a smaller number of PSNEs.

In $2$-player games we gave an exact formula for the frequency of games with a given number of PSNEs, and highlight that for less than $10$ strategies, games with a unique PSNE are more common than convergent games with multiple PSNEs, otherwise less common.

We believe that our graph-theoretic approach can be generally very useful to understand complicated games. Extensions of this work would include finding the analytical frequency of multi-player games with multiple pure Nash equilibria or with mixed Nash equilibria.

\clearpage

\section{Proofs}

\begin{proofof}{Theorem \ref{maintheorem}}
Consider the full $n$-partite graph for an $n$-player, $m$-strategy game. We order the nodes in the following way: $s_{-i}<s_{-j}$ for different players $i$ and $j$, if and only if $i<j$, and for the same player $i$, $s_{-i}<s'_{-i}$ under lexicographical ordering. Denote this full $n$-partite graph by $G^\text{f}=(V^\text{f},E^\text{f})$, where $V$ is the set of vertices and $E$ is the set of edges.

The Laplacian matrix of a graph $G=(V,E)$ without multiple edges and self-loops is defined as the square matrix with side length $|V|$ and
\begin{equation*}
\left(L(G)\right)_{ij} = \begin{cases}
\delta^+(i) & \text{if } i = j \\
-1 & \text{if } i\neq j, \,(i,j)\in E \\
0 & \text{if } i\neq j, \,(i,j)\not\in E
\end{cases}
\end{equation*}
where $\delta^+(i)$ is the out-degree of a node $i$. For $G^\text{f}$ described above, the Laplacian matrix takes the following form:
\begin{equation*}
\renewcommand*{\arraystretch}{1.5}
L\left(G^\text{f}\right) =
\begin{bmatrix}[columns-width = 0.7cm]
\setlength{\extrarowheight}{1mm}
D      & N_1    & 0      & \cdots & 0       \\
0      & \ddots & \ddots & \ddots & \Vdots  \\
\vdots & \ddots & \ddots & \ddots & 0       \\
0      &        & \ddots & \ddots & N_{n-1} \\
S      & 0      & \cdots & 0      & D
\end{bmatrix}
\end{equation*}
where $D,S,N_1,\dots,N_{n-1}$ are square matrices with side length $m^{n-1}$ defined as follows:
\begin{itemize}
\item $D=\text{diag}(m)$ is a diagonal matrix with $m$'s on the diagonal
\item $N_k=\text{diag}\left(K_1,\dots,K_{m^{k-1}}\right)$ is a blockmatrix with blockmatrices $K_l$ on the diagonal, where each $K_l$ has side length $m^{n-l}$ and consists of $m^2$ diagonal matrices $\text{diag}(-1)$, each with side length $m^{n-l-1}$.
\item $S$ is more irregular,
\begin{equation*}
(S)_{ij} = \begin{cases}
-1 & \text{if } \left(i \mod m^{n-2}\right) = \left\lfloor \frac{j-1}{m}\right\rfloor \\
0 & \text{otherwise}.
\end{cases}
\end{equation*}
\end{itemize}

For instance, in the case of $3$-player, $2$-strategy games, the Laplacian matrix corresponding to Figure \ref{fig:fc32} (left) is given by
\begin{equation*}
\renewcommand*{\arraystretch}{1}
L\left(G^\text{f}\right) = \left[
\begin{array}{cccc|cccc|cccc}
2 & 0 & 0 & 0  & -1 & 0  & -1 & 0 & 0 & 0 & 0 & 0 \\
0 & 2 & 0 & 0 & 0  & -1 & 0  & -1 & 0 & 0 & 0 & 0  \\
0 & 0 & 2 & 0  & -1 & 0  & -1 & 0 & 0 & 0 & 0 & 0 \\
0 & 0 & 0 & 2 & 0  & -1 & 0  & -1 & 0 & 0 & 0 & 0 \\ \hline
0 & 0 & 0 & 0 & 2 & 0 & 0 & 0  & -1  & -1 & 0 & 0 \\
0 & 0 & 0 & 0 & 0 & 2 & 0 & 0  & -1  & -1 & 0 & 0 \\
0 & 0 & 0 & 0 & 0 & 0 & 2 & 0 & 0 & 0  & -1  & -1 \\
0 & 0 & 0 & 0 & 0 & 0 & 0 & 2 & 0 & 0  & -1  & -1 \\ \hline
-1  & -1 & 0 & 0 & 0 & 0 & 0 & 0 & 2 & 0 & 0 & 0 \\
0 & 0  & -1  & -1 & 0 & 0 & 0 & 0 & 0 & 2 & 0 & 0 \\
-1  & -1 & 0 & 0 & 0 & 0 & 0 & 0 & 0 & 0 & 2 & 0 \\
0 & 0  & -1  & -1 & 0 & 0 & 0 & 0 & 0 & 0 & 0 & 2
\end{array}\right]
\end{equation*}
For a general blockmatrix $K=\left(\begin{matrix} A & B \\ C & D\end{matrix}\right)$, provided that $A$ is invertible, we have
\begin{equation*}
\det K = \det\left(D-C A^{-1} B\right) \det A.
\end{equation*}
Applying this identity iteratively to $L\left(G^\text{f}\right)$ yields

\begin{equation*}
\det L\left(G^\text{f}\right) = m^{m^{n-1}(n-1)} \cdot \det\left(D - \frac{1}{m^{n-1}}\cdot S\cdot \prod_{i=1}^{n-1} N_i\right).
\end{equation*}
There are $m^n$ ways to choose the first PSNE, each fixing $n$ nodes. Without loss of generality, we choose the nodes where each player chooses their first strategy. We condense these $n$ nodes to a single node representing the PSNE, see Figure \ref{fig:fc32}. The PSNE-node has an in-degree of $n(m-1)$; we delete all outgoing edges. The resulting $(n+1)$-partite graph consists of $nm^{n-1}-(n-1)$ nodes and will be denoted by $G^{\text{c}}=(V^\text{c},E^\text{c})$, where $V^\text{c}$ is the set of vertices and $E^\text{c}$ is the set of edges. All nodes except the PSNE-node are free.

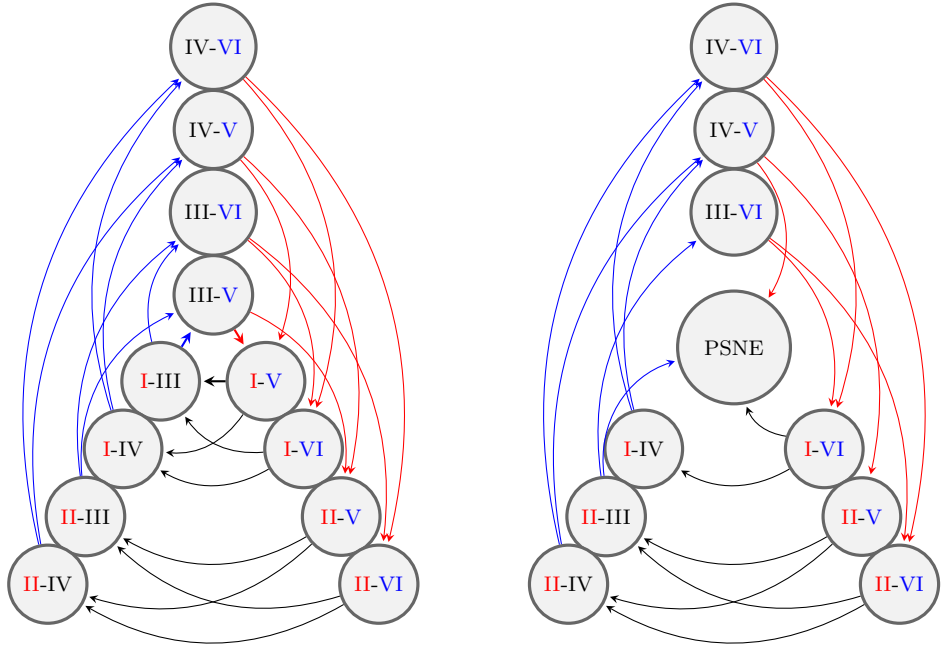
\begin{figure}[H]
\begin{minipage}{0.495\textwidth}
\centering
\normalfont
\begin{tikzpicture}%
  [>=stealth,
   shorten >=1pt,
   node distance=1.1cm,
   on grid,
   auto,
   every state/.style={draw=black!60, fill=black!5, very thick, minimum width = 0.9cm}
  ]
\node[state] (A1)               {\scriptsize IV-\textcolor{blue}{VI}};
\node[state] (A2) [below=of A1] {\scriptsize IV-\textcolor{blue}{V}}; 
\node[state] (A3) [below=of A2] {\scriptsize III-\textcolor{blue}{VI}};
\node[state] (A4) [below=of A3] {\scriptsize III-\textcolor{blue}{V}};

\node[state] (B1) [below right=1.15cm and 0.7cm of A4] {\scriptsize \textcolor{red}{I}-\textcolor{blue}{V}};
\node[state] (B2) [below right=0.9cm and 0.5cm of B1] {\scriptsize \textcolor{red}{I}-\textcolor{blue}{VI}};
\node[state] (B3) [below right=0.9cm and 0.5cm of B2] {\scriptsize \textcolor{red}{II}-\textcolor{blue}{V}};
\node[state] (B4) [below right=0.9cm and 0.5cm of B3] {\scriptsize \textcolor{red}{II}-\textcolor{blue}{VI}};

\node[state] (C1) [below left=1.15cm and 0.7cm of A4] {\scriptsize \textcolor{red}{I}-III};
\node[state] (C2) [below left=0.9cm and 0.5cm of C1] {\scriptsize \textcolor{red}{I}-IV};
\node[state] (C3) [below left=0.9cm and 0.5cm of C2] {\scriptsize \textcolor{red}{II}-III};
\node[state] (C4) [below left=0.9cm and 0.5cm of C3] {\scriptsize \textcolor{red}{II}-IV};

\path[->]
(A1) edge[red, bend left=30] node {} (B2)
(A1) edge[red, bend left=30] node {} (B4)
(A2) edge[red, bend left=30] node {} (B1)
(A2) edge[red, bend left=30] node {} (B3)
(A3) edge[red, bend left=30] node {} (B2)
(A3) edge[red, bend left=30] node {} (B4)
(A4) edge[red, bend left=00, line width=0.8pt] node {} (B1)
(A4) edge[red, bend left=35] node {} (B3)

(B1) edge[black, bend left=0, line width=0.8pt] node {} (C1)
(B1) edge[black, bend left=30] node {} (C2)
(B2) edge[black, bend left=30] node {} (C1)
(B2) edge[black, bend left=30] node {} (C2)
(B3) edge[black, bend left=30] node {} (C3)
(B3) edge[black, bend left=30] node {} (C4)
(B4) edge[black, bend left=30] node {} (C3)
(B4) edge[black, bend left=30] node {} (C4)

(C1) edge[blue, bend left=30] node {} (A3)
(C1) edge[blue, bend left=0, line width=0.8pt] node {} (A4)
(C2) edge[blue, bend left=30] node {} (A1)
(C2) edge[blue, bend left=30] node {} (A2)
(C3) edge[blue, bend left=30] node {} (A3)
(C3) edge[blue, bend left=35] node {} (A4)
(C4) edge[blue, bend left=30] node {} (A1)
(C4) edge[blue, bend left=30] node {} (A2)
;
\end{tikzpicture}
\end{minipage}
\hfill
\begin{minipage}{0.495\textwidth}
\centering
\normalfont
\begin{tikzpicture}%
  [>=stealth,
   shorten >=1pt,
   node distance=1.1cm,
   on grid,
   auto,
   every state/.style={draw=black!60, fill=black!5, very thick, minimum width = 0.9cm}
  ]
\node[state] (A1)               {\scriptsize IV-\textcolor{blue}{VI}};
\node[state] (A2) [below=of A1] {\scriptsize IV-\textcolor{blue}{V}}; 
\node[state] (A3) [below=of A2] {\scriptsize III-\textcolor{blue}{VI}};
\node[state, transparent] (A4) [below=of A3] {\scriptsize III-\textcolor{blue}{V}};

\node[state, transparent] (B1) [below right=1.15cm and 0.7cm of A4] {\scriptsize \textcolor{red}{I}-\textcolor{blue}{V}};
\node[state] (B2) [below right=0.9cm and 0.5cm of B1] {\scriptsize \textcolor{red}{I}-\textcolor{blue}{VI}};
\node[state] (B3) [below right=0.9cm and 0.5cm of B2] {\scriptsize \textcolor{red}{II}-\textcolor{blue}{V}};
\node[state] (B4) [below right=0.9cm and 0.5cm of B3] {\scriptsize \textcolor{red}{II}-\textcolor{blue}{VI}};

\node[state, transparent] (C1) [below left=1.15cm and 0.7cm of A4] {\scriptsize \textcolor{red}{I}-III};
\node[state] (C2) [below left=0.9cm and 0.5cm of C1] {\scriptsize \textcolor{red}{I}-IV};
\node[state] (C3) [below left=0.9cm and 0.5cm of C2] {\scriptsize \textcolor{red}{II}-III};
\node[state] (C4) [below left=0.9cm and 0.5cm of C3] {\scriptsize \textcolor{red}{II}-IV};

\node[state, minimum size=1.5cm] (A) [below=0.7cm of A4] {\scriptsize PSNE};

\path[->]
(A1) edge[red, bend left=30] node {} (B2)
(A1) edge[red, bend left=30] node {} (B4)
(A2) edge[red, bend left=35] node {} (A)
(A2) edge[red, bend left=30] node {} (B3)
(A3) edge[red, bend left=30] node {} (B2)
(A3) edge[red, bend left=30] node {} (B4)

(B2) edge[black, bend left=30] node {} (A)
(B2) edge[black, bend left=30] node {} (C2)
(B3) edge[black, bend left=30] node {} (C3)
(B3) edge[black, bend left=30] node {} (C4)
(B4) edge[black, bend left=30] node {} (C3)
(B4) edge[black, bend left=30] node {} (C4)

(C2) edge[blue, bend left=30] node {} (A1)
(C2) edge[blue, bend left=30] node {} (A2)
(C3) edge[blue, bend left=30] node {} (A3)
(C3) edge[blue, bend left=40] node {} (A)
(C4) edge[blue, bend left=30] node {} (A1)
(C4) edge[blue, bend left=30] node {} (A2)
;
\end{tikzpicture}
\end{minipage}
\caption{For $3$-player, $2$-strategy games the full graph $G^\text{f}$ on the left and the condensed graph $G^\text{c}$ on the right.}
\label{fig:fc32}
\end{figure}

We apply Kirchhoffs theorem to $G^{\text{c}}$ to get the number of spanning trees. This guarantees that the game converges under clockwork best-response dynamics. Kirchhoffs theorem (applied to our problem) states that the number of spanning trees is the determinant of the Laplacian matrix of $G^\text{c}$ with the first row and column deleted, which corresponds to the PSNE-node.

For a quadratic matrix A with side length $n$, we define $\widetilde{A}$ to be the quadratic matrix with side length $(n-1)$ obtained from $A$ by deleting the first row and column. We can compute $\det \widetilde{L\left(G^\text{c}\right)}$ by modifying the formula for $\det L\left(G^\text{f}\right)$, namely
\begin{equation*}
\det \widetilde{L\left(G^\text{c}\right)} = m^{\left(m^{n-1}-1\right)(n-1)} \cdot \det\left(\widetilde{D} - \frac{1}{m^{n-1}}\cdot \widetilde{S}\cdot \prod_{i=1}^{n-1} \widetilde{N_i}\right).
\end{equation*}
The matrix $\widetilde{S}\cdot\prod_i \widetilde{N_i}$ is given by
\begin{equation*}
\left(\widetilde{S}\cdot\prod_i \widetilde{N_i}\right)_{ij} = m - \mathbbm{1}_{\left[1,m^{n-\delta(i)}-1\right]}(j)
\end{equation*}
where
\begin{equation*}
\delta(i) := \argmin_{p\in [1,n-1]} \left(\min_{k\in [1,m^p]}\left(\left|i-km^{n-p-1}\right|\right)\right).
\end{equation*}
We simplify the matrix $\widetilde{D} - \frac{1}{m^{n-1}}\cdot \widetilde{S}\cdot \prod_{i=1}^{n-1} \widetilde{N_i}$ by elementary row- and column-operations to obtain a matrix $A$ by the following algorithm:
\begin{algorithm}[H]
\caption{Simplifying $\widetilde{D} - \frac{1}{m^{n-1}}\cdot \widetilde{S}\cdot \prod_{i=1}^{n-1} \widetilde{N_i}$ to obtain $A$}

\begin{enumerate}
\item For $p \in [1,\dots,n-1]$:
\begin{enumerate}
\item For $i\in [1,\dots,m^{n-1}-1]$:
\begin{enumerate}
\item If $i=m^{p-1}$ or $\delta(i)\neq n-p$, continue.
\item Subtract the $m^{p-1}$'s row from $i$.
\end{enumerate}
\end{enumerate}
\item For $k \in [1,\dots,n^{m-1}-1]$:
\begin{enumerate}
\item If for any $p\in [0,\dots,n-1]$, $k|m^p$, continue.
\item Add column $k$ to column $m^{n-\delta(k)-1}$.
\end{enumerate}
\end{enumerate}
\end{algorithm}
\noindent The determinant of the matrix $A$ can be written as
\begin{equation*}
\det A = m^{m^{m-1}-n} \cdot\det \widehat{A}
\end{equation*}
for a matrix $\widehat{A}$ with side length $(n-1)$ and given by
\begin{equation*}
\left(\widehat{A}\right)_{ij} = \begin{cases}
m+\frac{m-1}{m^{n-1}}-\frac{m-1}{m^{i-1}} & i=j \\
\frac{m-1}{m^{n-i+j-1}}-\frac{m-1}{m^{j-1}} & i>j \\
-\frac{m-1}{m^{j-1}} & i<j
\end{cases}
\end{equation*}
Adding the $i$-th column multiplied by $\left(-\frac{1}{m}\right)$ to the $(i+1)$-th column for $i=n-2,n-3,\dots,1$, we get a matrix of the following form
\begin{equation*}
\renewcommand*{\arraystretch}{1.5}
\begin{bmatrix}[columns-width = 0.7cm]
\setlength{\extrarowheight}{1mm}
D_1      & N    & 0      & \cdots & 0       \\
E_2     & D & \ddots & \ddots & \Vdots  \\
\vdots & 0 & \ddots & \ddots & 0       \\
\vdots &   \vdots     & \ddots & \ddots & N \\
E_{n-1}      & 0      & \cdots & 0      & D
\end{bmatrix}
\end{equation*}
where
\begin{align*}
D_1 ={}& \frac{m-1}{m^{n-1}}+1 \\
E_i ={}& \frac{m-1}{m^{n-i}} - (m-1) \\
D ={}& m \\
N ={}& -\frac{m-1}{m^n}-1
\end{align*}
To eliminate the $N$-entries on the upper diagonal, we add the $i$-th row multiplied by
\begin{equation*}
F := -\frac{N}{D} = \frac{\frac{m-1}{m^n}+1}{m}
\end{equation*}
to the $(i-1)$-th row for $i=n-1,\dots,2$. Then the matrix is lower-triangular and the $D_1$-entry is given by
\begin{align*}
\widetilde{D_1} ={}& D_1+F\cdot E_2+\cdots+F^{n-2}\cdot E_{n-1} \\
={}& D_1+\sum_{i=0}^{n-3} F^{i+1}\cdot E_{i+2} \\
={}& D_1+ \sum_{i=0}^{n-3} F^{i+1} \left(\frac{m-1}{m^{n-i-2}}\right) - \sum_{i=0}^{n-3} F^{i+1} (m-1) \\
={}& D_1 + F\left(\frac{m-1}{m^{n-2}}\right) \sum_{i=0}^{n-3} \left(Fm\right)^i - F(m-1) \sum_{i=0}^{n-3} F^i \\
={}& D_1 + F\left(\frac{m-1}{m^{n-2}}\right) \left(\frac{(Fm)^{n-2}-1}{Fm-1}\right) - F(m-1)\left(\frac{F^{n-2}-1}{F-1}\right) \\
={}& m(r-1)+m\left(r^{n-1}-r\right)-\frac{r(m-1)}{r-m} \left(\left(\frac{r}{m}\right)^{n-2}-1\right)+1
\end{align*}
where $r:=\frac{m-1}{m^n}+1$, and then
\begin{align*}
\det \widehat{A} ={}& m^{n-2}\cdot\widetilde{D_1}.
\end{align*}
Finally, the frequency of games with exactly one PSNE is given by
\begin{equation*}
p^1_{n,m} = \frac{m^n}{m^{nm^{n-1}}} \det \widetilde{L\left(G^\text{c}\right)} = \frac{1}{m^{m^{n-1}-1}} \det A = \frac{1}{m^{n-1}} \det \widehat{A} = \frac{1}{m}\widetilde{D_1}
\end{equation*}
where we have multiplied by the number of possible positions of the PSNE and divided by the total number of possible arrangements. This completes the proof.
\end{proofof}

\begin{proofof}{Theorem \ref{upperbound}}
Consider a full $n$-partite graph and assign $k$ PSNEs, thereby fixing the outgoing edges of $k n$ nodes. We show that the number of possible realizations as a game decreases, when adding another PSNE.

The number of ways we can add another PSNE (which is, in general, very complicated to compute) is bounded from above by $\left(m^{n-1}-k\right)m=m^n-km$, which is because there are $m^{n-1}-k$ free nodes for each player, each free node has an out-degree of $m$, and fixing two nodes of an $n$-cycle fixes the remaining ones. However, adding a PSNE decreases the number of possible realizations as a game by a factor of $m^n-1$, because the $n$ nodes may not form a cycle.

Induction over the number of added PSNEs completes the proof.
\end{proofof}

\begin{proofof}{Theorem \ref{twoplayers}}
It was shown in Austin \cite{TLA} that the number of chromatic digraphs with $m$ nodes of each type, where each node has an out-degree one, and with a cycle of length $2k$, $1\leq k\leq m$, is
\begin{equation*}
(2m-k) \left(m^{m-k-1}\right)^2 \left(\frac{m!}{(m-k)!}\right)^2.
\end{equation*}
Factoring out the number of ways to arrange $k$ vertices on a cycle ($(k+1)!$) and the total number of possible arrangements ($m^{2m})$, we get
\begin{equation*}
p_{2,m}^k = \frac{2m-k}{m^{2k+2} (k-1)!}  \left(\frac{m!}{(m-k)!}\right)^2.
\end{equation*}
This was given in Pangallo et al. \cite{MP} as a recursively defined formula.
\end{proofof}

\vfill

\bibliographystyle{elsarticle-num-names}

\end{document}